\documentclass[useAMS,usenatbib]{mn2e}
\usepackage{psfrag,graphicx}
\usepackage{color}
\usepackage{txfonts}
\usepackage{breqn}
\usepackage{graphicx}
\title[Black hole formation in the early universe]
   {Black hole formation in the early universe}

\author[Latif et al.]
  {M.~A.~Latif,$^1$
  D.~R.~G.~Schleicher,$^1$ 
  W.~Schmidt,$^1$
  J.~Niemeyer$^1$
   \newauthor % starts a new line in the
   $^1$ Institut f\"ur Astrophysik, Georg-August-Universit\"at, \\
    Friedrich-Hund-Platz 1, D-37077 G\"ottingen, Germany}
\date{today}

\pagerange{\pageref{firstpage}--\pageref{lastpage}} \pubyear{2009}

\def\LaTeX{L\kern-.36em\raise.3ex\hbox{a}\kern-.15em
      T\kern-.1667em\lower.7ex\hbox{E}\kern-.125emX}

\begin{document}

\bibliographystyle{mn2e}

\label{firstpage}

 \maketitle
% 

% \newcommand{\newln}{\\&\quad\quad{}}
% \date{today}

\begin{abstract}
{Supermassive black holes with up to a $\rm 10^{9}~M_{\odot}$ dwell in the centers of present-day galaxies, and their presence has been confirmed at z $\geq$ 6. Their formation at such early epochs is still an enigma. Different pathways have been suggested to assemble supermassive black holes in the first billion years after the Big Bang. Direct collapse has emerged as a highly plausible scenario to form black holes as it provides seed masses of $\rm 10^{5}-10^{6}~M_{\odot}$. Gravitational collapse in atomic cooling haloes with virial temperatures T$_{vir} \geq 10^{4}$~K may lead to the formation of massive seed black holes in the presence of an intense background UV flux. Turbulence plays a central role in regulating accretion and transporting angular momentum. We present here the highest resolution cosmological large-eddy simulations to date which track the evolution of high-density regions on scales of $0.25$~AU beyond the formation of the first peak, and study the impact of subgrid-scale turbulence. The peak density reached in these simulations is $\rm 1.2 \times 10^{-8}~g~cm^{-3}$. Our findings show that while fragmentation occasionally occurs, it does not prevent the growth of a central massive object resulting from turbulent accretion and occasional mergers. The central object reaches $\rm \sim 1000~M_{\odot}$ within $4$ free-fall times, and we expect further growth up to $\rm 10^{6}~M_{\odot}$ through accretion in about 1 million years. The direct collapse model thus provides a viable pathway of forming high-mass black holes at early cosmic times.}
 
\end{abstract}
% With accretion at the Eddington limit, obtaining such masses during the available time is very difficult when starting from a stellar black hole. The pathways towards the formation of supermassive black holes have already been pointed out including the direct collapse of massive gas clouds, but also of a dense star cluster due to relativistic instability., possibly supermassive star as an intermediate stage% context

% aims

%  results

% conclusion 


\section{Introduction}
The formation of pregalactic black holes is a fundamental issue in the study of galaxy formation. We know that most galaxies have black holes in their nuclei. Numerous quasars with black hole masses of $\rm ~10^{9}~M_{\odot}$ have been observed at redshifts higher than six when the Universe was less than a billion years old \citep{2003AJ....125.1649F,2006AJ....131.1203F}. Recently, the discovery of a z=7.1 quasar with a black hole mass of $\rm ~ 2 \times 10^{9}~M_{\odot}$ is even more striking \citep{2011Natur.474..616M}. It is however not very well understood how these quasars observed in the Sloan Digital Sky Survey could attain such high masses in a such short span of time after the Big Bang. It suggests that black hole seeds have been formed even earlier and it is important to get a physical understanding of their origin.

One scenario for the formation of supermassive black holes (SMBHs) could be that they formed through accretion and merging of stellar remnants \citep{2004gimi.confE..14H, 2004ApJ...613...36H}. Numerical studies show low gas densities in the vicinity of black holes remnants due to their feedback which brings an obstruction for their formation though accretion and merging \citep{2007MNRAS.374.1557J, 2008arXiv0811.0820A}. Black holes may have formed through runaway collisions of young dense stellar clusters due to relativistic instability \citep{2004Natur.428..724P,2008ApJ...686..801O, 2009ApJ...694..302D} but this can happen only if the Universe is enriched with trace amounts of metals after the formation of the first stars between redshift 10-20, and the resulting black hole masses are about $\rm 10^{3}~M_{\odot}$. Primordial black holes may have formed during the early stages of the Big Bang albeit we do not see any observational evidence of their formation \citep{2008ApJ...680..829R}. Various other ways for the formation of massive black holes has been suggested \citep{1978Obs....98..210R,2008arXiv0803.2862D,2009ApJ...696.1798T,2009ApJ...702L...5B,2009MNRAS.396..343R,2010A&ARv..18..279V,2012ApJ...750...66J,2012arXiv1203.6075H,2013arXiv1301.5567P,2013ApJ...764...72M,2013arXiv1304.1369C,2013arXiv1304.4601J,2013arXiv1304.7787R}. Protogalactic metal free haloes may directly collapse to form supermassive black holes provided that the fragmentation in these halos remains suppressed \citep{2002ApJ...569..558O,2003ApJ...596...34B,2006ApJ...652..902S,2006MNRAS.370..289B,2008MNRAS.391.1961D,2008arXiv0803.2862D,2010MNRAS.402.1249S,2010ApJ...712L..69S,2011MNRAS.411.1659L}. Given the difficulties with other methods for the formation of supermassive black holes, direct collapse emerges as a plausible scenario. Haloes with virial temperatures $\rm \geq 10^{4}$ K are the most likely candidates for the formation of such objects. The formation of black holes via direct collapse is feasible if fragmentation is sufficiently suppressed, so that most of the material is accreted onto the central object.

The essential condition for the direct collapse scenarios is to avoid fragmentation. Molecular hydrogen is the only coolant in metal free haloes which can bring the temperature of the gas down to a few hundred Kelvins and may induce fragmentation by reducing the Jeans mass. The gas is unlikely to fragment to form stars in primordial halos in the absence of H$_{2}$ cooling \citep{2003ApJ...592..975L}. For zero metallicity atomic cooling haloes devoid of molecules, the Lyman alpha radiation is the only pathway to cool the gas down to 8000 K. Cosmological simulations performed to study the collapse of protogalactic metal free halos  show that self gravitating gas collapses almost isothermally in the absence of H$_{2}$ cooling  \citep{2003ApJ...596...34B,2008ApJ...682..745W,2009MNRAS.393..858R,2011MNRAS.411.1659L}.
% Intense ultraviolet (UV) flux photo-dissociates $\rm H_{2}$ molecules in atomic cooling haloes and may justify the assumption for the insignificance of $\rm H_{2}$ molecules in these haloes.

In this respect,the presence of UV radiation has important implications for the formation of supermassive black holes as it may suppress cooling and consequently halt the fragmentation of a gas cloud. It is found that strong Lyman Werner flux $\rm \geq 10^{3}$ in units of $\rm J_{21}$  is required to photo-dissociate $\rm H_{2}$ molecules in haloes with virial temperatures $\rm >10^{4}$ K \citep{2001ApJ...546..635O,2007MNRAS.374.1557J,2008MNRAS.391.1961D,2010MNRAS.402.1249S,2010ApJ...712L..69S,2011MNRAS.418..838W,2011A&A...532A..66L,2012MNRAS.425.2854A,2013MNRAS.tmp..551L}.

Turbulent energy plays a key role in the formation of structures as it locally compresses the gas to high densities where it may become susceptible to fragmentation and collapse. It can provide an additional pressure against the gravitational collapse at larger scales. In order to transport the angular momentum, previous studies suggested scenarios based on gravitational instabilities in self-gravitating disks  \citep{2004MNRAS.354..292K} and self-regulation due to the presence of turbulence \citep{2009ApJ...702L...5B}. In a similar fashion, turbulence is thought to regulate the angular momentum transport in minihalos \citep{2002Sci...295...93A,2012ApJ...745..154T}. In the presence of high mass accretion rates and efficient cooling, gravitational torques may not be sufficient to transport the gas through the disk leading to fragmentation \citep{Clark11,Greif12}. Numerical simulations show that primordial gas becomes turbulent during the formation of the first galaxies \citep{2008ApJ...682..745W,2008MNRAS.387.1021G}. While these studies employed a typical Jeans length resolution of 16 cells, later studies demonstrated the necessity of a higher Jeans resolution i.e., minimum resolution of 32 cells per Jeans length to obtain converged turbulent energy \citep{2011ApJ...731...62F, 2012ApJ...745..154T,2013MNRAS.tmp..551L}.  
In addition, the turbulent energy cascade is expected to generate turbulence even on subgrid-scales (SGS), which may backreact on the resolved scales via turbulent viscosity. \cite{2013MNRAS.tmp..551L} have shown that applying a SGS turbulence model to cosmological AMR simulations significantly changes the morphology of the halo during its initial collapse phase in comparison to standard simulations without an explicit SGS model. This may have a significant impact on the the formation of astrophysical objects.

The main objective of this work is to assess the role of turbulence in the formation of seed black holes. To accomplish this goal, we have performed the highest resolution cosmological large eddy simulations to date which track the gravitational collapse beyond the formation of the first peak in the presence of a strong Lyman Werner background UV flux. We employ a fixed Jeans length of 64 cells throughout the evolution of simulations to resolve turbulent eddies and follow the collapse down to sub AU scales for 4 free-fall times by exploiting the adaptive mesh technique. This study will allow us to explore the feasibility of the direct collapse scenario. It will further enable us to quantify the impact of subgrid/resolved turbulence in the formation of supermassive black holes.

Our paper is organized as follows. In the next section, we describe the simulation setup and the numerical methods employed. In the 3rd section of the paper, we present the results obtained in this study. In the last section of this article, we summarize our main results and confer our conclusions.

\section{Basic setup}

\subsection{Computational methods}

We present high resolution cosmological simulations which capture a large range of spatial scales starting from Mpc down to sub AU scales. These simulations were performed using a modified version of the adaptive mesh refinement code Enzo \citep{2004astro.ph..3044O} including a subgrid-scale (SGS) model which takes into account unresolved turbulence \citep{SchmNie06b}. Enzo is an adaptive mesh refinement (AMR), parallel, grid-based cosmological hydrodynamics code \citep{2004astro.ph..3044O,2007arXiv0705.1556N}. It makes use of the message passing interface (MPI) to achieve portability and scalability on parallel systems. The computational domain is discretized using a nested grid structure. We use a split hydro solver with a 3rd order piece-wise parabolic (PPM) method for hydrodynamical calculations. The dark matter N-body dynamics is solved using the particle-mesh technique. A multigrid Poisson solver is employed for the self-gravity computations.

The simulations are commenced with cosmological initial conditions generated from Gaussian random fields. We make use of the inits package available with Enzo to create nested grid initial conditions. Our simulations start at redshift $\rm z=100$ with a top grid resolution of $\rm 128^{3}$ cells and we select a massive halo at redshift 15 using the halo finder  based on a standard friends of friends algorithm \citep{2011ApJS..192....9T}. Our computational volume has a cosmological size of 1 Mpc $\rm h^{-1}$ and is centered on the most massive halo. Two additional nested levels of refinement are subsequently employed each with a resolution of $\rm 128^{3}$ cells.  In all, we initialize 5767168  particles to compute the evolution of the dark matter dynamics yielding a dark matter resolution of 620 $\rm M_{\odot}$. The parameters from the WMAP seven years data \citep{2011ApJS..192...14J} are used for creating the initial conditions. We further allow additional 27 levels of refinement in the central 62 kpc region of the halo during the course of simulation yielding a physical resolution of sub AU scales (3 AU in comoving units). The resolution criteria used in these simulations are based on the Jeans length, the gas over-density and the particle mass resolution. The gas overdensity criterion flags grid cells for refinement when the gas density exceeds four times the mean baryonic density. Similarly, grids are marked for refinement when the dark matter density exceeds 0.0625 times  $ \rho_{DM}r^{l \alpha}$ where r=2 is the refinement factor, l is the refinement level and $\alpha =-0.3$ makes the refinement super-Lagrangian. The grid cells matching these requirement are marked for refinement. We mandated a Jeans length resolution of 64 cells throughout the evolution. This criterion was applied during the course of the simulations to ensure a high resolution in the central collapsing region, simultaneously ensuring the Truelove criterion \citep{1997ApJ...489L.179T}. When the highest refinement level is reached, the thermal evolution becomes adiabatic. This approach enables us to follow the evolution of structures beyond the formation of the first peak until they reach a peak density of $\rm 1.2 \times 10^{-8}~ g/cm^{3}$ which corresponds to four free-fall times. Our study consists of 9 distinct halos selected from various Gaussian random seeds. The results from the SGS turbulence model are compared with normal runs. In total, we perform 18 simulations.

%  the particle mass resolution criterion ensures that the minimum mass of dark matter particles is refined during the simulation.

\subsection{Chemistry}

According to Big Bang nucleosynthesis, primordial gas is mainly composed of hydrogen and helium. For zero metallicity gas devoid of molecules, atomic line cooling is the only pathway to cool the gas down to 8000 K. While molecular hydrogen can potentially cool below a few thousand K, the presence of strong UV flux emitted by the first generation of stars has important implications for structure formation \citep{2001ApJ...546..635O,2008ApJ...686..801O,2011MNRAS.418..838W}. A photon with energy between 11.2 and 13.6 eV  is absorbed in the Lyman Werner bands of molecular hydrogen and brings it into an excited state. The $\rm H_{2}$ molecule then decays to the vibrational continuum of a ground state and is dissociated. This two step process is known as the Solomon process \citep{1997NewA....2..181A}. The photo-dissociation reaction is given by
\begin{equation}
H_{2} + \gamma  \rightarrow  H_{2}^{*} \rightarrow H + H .
\label{photo}
\end{equation}
The critical value of the flux can be obtained by balancing the $\rm H_{2}$ formation and dissociation time scales. The value of the critical UV intensity is higher for the larger halos as compared to the minihalos and varies from halo to halo. It is found that a strong Lyman Werner flux $\rm \geq 10^{3}$ in units of $\rm J_{21}$ is required to photo-dissociate $\rm H_{2}$ molecules in halos with virial temperatures $>10^{4}$ K for a radiation temperature of $\rm 10^{5}$ K \citep{2001ApJ...546..635O,2007MNRAS.374.1557J,2008MNRAS.391.1961D,2010ApJ...712L..69S,2011MNRAS.418..838W,2011A&A...532A..66L,2012MNRAS.426.1159S}. $\rm J_{21}$ is the background UV flux below the Lyman limit (i.e. 13.6 eV) in units of $\rm 10^{-21}~erg~cm^{-2}~s^{-1}~Hz^{-1}~sr^{-1}$.

% The $\rm H_{2}$ can be photo-dissociated by the photons with energy between 11.2 and 13.6 eV (Lyman Werner photons). For a stellar spectrum of $10^{5}$~K, a photon with this energy is absorbed in the Lyman Werner bands of molecular hydrogen and puts it in excited state. The $\rm H_{2}$ molecule then decays to the vibrational continuum of a ground state and is dissociated. This two step process is known as the Solomon process. The photo-dissociation reaction is given by
% 
% \begin{equation}
% H_{2} + \gamma  \rightarrow  H_{2}^{*} \rightarrow H + H .
% \label{photo}
% \end{equation}
% The critical value of flux can be obtained by balancing the $\rm H_{2}$ formation and dissociation time scales. The value of the critical UV intensity is higher for the larger haloes as compared to the minihaloes and varies from a halo to halo.

To study the thermal evolution of the gas, it is crucial to model the chemistry of primordial gas in detail. In order to include the primordial non-equilibrium chemistry, the rate equations of $\rm H$, $\rm H^{+}$,$\rm He$, $\rm He^{+}$,~$\rm He^{++}$, $\rm e^{-}$,~$\rm H^{-}$,~$\rm H_{2}$,~$\rm H_{2}^{+}$ are self-consistently solved in the cosmological simulations. We use the $\rm H_{2}$ photo-dissociating background UV flux implemented in the Enzo code, and employ an external UV field of constant strength $\rm 10^{3}$ in units of $\rm J_{21}$. We presume that such flux is generated from a nearby star forming halo \citep{2008MNRAS.391.1961D} and is emitted by Pop III stars with a thermal spectrum of $\rm 10^{5}$ K. We include all relevant cooling and heating mechanisms like collisional ionization cooling, radiative recombination cooling, collisional excitation cooling, H$_{2}$ cooling as well as $\rm H_{2}$ formation heating. The chemistry solver used in this work is a modified version of \cite{1997NewA....2..181A} and \cite{1997NewA....2..209A} (see \cite{2012ApJ...745..154T} for details).

We note that for an optical depth $\tau_{0} > 10^{7}$, the escape time of Lyman alpha photons becomes longer than the gas free fall time. In fact, \cite{2006ApJ...652..902S} suggested that at columns above $\rm 10^{22}~ cm^{-2}$ Lyman alpha photons get trapped inside the halo which suppresses the cooling and may lead to an adiabatic collapse. The impact of Lyman alpha trapping was explored in detail by \cite{2001ApJ...546..635O,2010ApJ...712L..69S}, finding that cooling due to additional processes becomes relevant in these cases particularly the two-photon decay (2s-1s transition) and the 3p-2s transition of atomic hydrogen. Effectively, the temperature evolution is then very close to the evolution obtained from optically thin Lyman $\alpha$ cooling. We also note that for a radiation temperature of $\rm 10^{5}$ K, $\rm H_{2}$ is mainly dissociated by the Solomon process while for the stellar spectrum of $\rm 10^{4}$ K the main dissociation route is $\rm H^{-}$ \citep{2010MNRAS.402.1249S,2011MNRAS.418..838W}. In principle, one would of course expect the presence of both contributions. As we focus on a situation where H$_{2}$ cooling is not relevant, the details of these processes are however not important here. We neglect here the effect of $\rm H_{2}$ self-shielding which could potentially raise the strength of the Lyman Werner flux required to photo-dissociate molecular hydrogen. The presence of X-rays/cosmic rays flux may boost the formation of $\rm H_{2}$ and further increase the threshold of UV flux \citep{2009ApJ...696.1798T,2011MNRAS.416.2748I}. 

\subsection{Turbulence Subgrid Scale Model}
The phenomenon of turbulence describes a process where energy cascades from large scales to small scales through a series of eddies. The formation and evolution of cosmic structures is a typical case of turbulence generation. Due to the high Reynolds numbers relevant for astrophysical systems, it is not possible to resolve all scales down to the dissipative scale even with adaptive mesh refinement techniques. In AMR simulations, mainly the core of the halo is well resolved and turbulence cascades from coarser grids corresponding to large scales down to the center of the halos without being properly accounted for.
% WS - don't repeat this: In engineering and many other disciplines of computational fluid dynamics SGS turbulence models are used to 
% represent the effect of unresolved turbulence on resolved scales. 
To compute the unresolved turbulence on grid scales in our simulation, we perform large eddy simulations (LES) with the SGS turbulence model proposed in \cite{SchmNie06b}. This SGS model is based on a mathematically rigorous approach separating the resolved and unresolved scales, and connecting them via an eddy-viscosity closure for the non-linear energy transfer across the grid scale. The turbulent viscosity is given by the grid scale and the SGS turbulence energy, i.~e., the kinetic energy associated with numerically unresolved turbulent velocity fluctuations. In contrast, implicit large eddy simulations (ILES) use only the numerical dissipation stemming from the
discretization errors of the compressible fluid dynamics equations. This is the standard method in computational astrophysics.

Applying the filtering mechanism to the fluid equations and solving them in a comoving coordinate system, we obtain \citep{SchmNie06b,2009ApJ...707...40M}:
% \begin{equation}
%  {\partial \over \partial t} \langle \tilde{\rho} \rangle + {1 \over a } { \partial \over \partial x_{j}} \hat{u_{j}} \langle \tilde{\rho} \rangle = 0 
% \end{equation}
% \begin{equation}
\begin{dmath}
{\partial \over \partial t} \rho + {1 \over a } { \partial \over \partial x_{j}} v_{j} \rho = 0 
\end{dmath}
\begin{dmath}
{ \partial \over \partial t} \rho v_{j} + {1 \over a }{\partial \over \partial x_{i}} v_{j} \rho v_{i} = -{1 \over a } {\partial \over \partial x_{j}} P + \rho
g_{j}^{*} -{1 \over a }{\partial \over \partial x_{j}} \tau_{ij}  -{\dot{a} \over a } \rho v_{j}
\end{dmath}
\begin{dmath}
{ \partial \over \partial t} \rho e_{\rm res} + {1 \over a }{\partial \over \partial x_{j}} v_{j} \rho e_{\rm res} =  -{1 \over a } {\partial \over \partial x_{i}} v_{i} P + {1 \over a }v_{i} \rho g_{i}^{*} - {\dot{a} \over a } (\rho e_{\rm res} + {1 \over 3 }  \rho v_{i} v_{i} + P) + 
 \rho (\lambda + \epsilon)-{1 \over a } v_{i} {\partial \over \partial x_{j}} \tau_{ij}
\end{dmath}
\begin{dmath}
{ \partial \over \partial t} \rho e_{\rm t} + {1 \over a } { \partial \over \partial x_{j}} v_{j} \rho e_{\rm t} = \mathbb{D} - \rho (\lambda + \epsilon)  - {1 \over a } \tau_{ij} {\partial \over \partial x_{j}} v_{i} - 2{\dot{a} \over a } \rho e_{\rm t}
\end{dmath}

Here $a(t)$ is the scale factor of the cosmological expansion, $\rho$ and $P$ are the comoving density and pressure, $v_{i}$ is the peculiar velocity, $g_i^*$ is the gravitational acceleration, $e_{\rm res}=e_{\rm int}+ \frac{1}{2}v^2$ the sum of the numerically resolved thermal and kinetic energies, and $e_{\rm t}$ the SGS turbulent energy (see \cite{SchmNie06b,2009ApJ...707...40M} for details). The terms $\mathbb{D}$, $\lambda$, $\epsilon$, and $\tau_{ij}$ correspond to SGS transport (diffusion), pressure dilatation, dissipation, and turbulent stresses. The following closure relations in terms of the resolved flow and the SGS turbulent velocity $q=\sqrt{2 e_{\rm t}}$ are applied to evaluate these terms:
% \ch{where}
\begin{eqnarray}
\mathbb{D} &=& { \partial \over \partial r_{i}} C_{\mathbb{D}} \rho  l_{\Delta}q^{2} { \partial \over \partial r_{i}} q,\\
\lambda &=& C_{\lambda}q^{2}{\partial \over \partial r_{i}} \nu_{i},\\
\epsilon &=& C_{\epsilon} {q^{3} \over l_{\Delta}},\\ %\left(1 +\alpha_{1}  M^{2}_{t} \right)
\tau_{ij} &=& -2 \eta_{t} S^{*}_{ij} + {1 \over 3}\delta_{ij} \rho q^{2},
\label{edis}
\end{eqnarray}
where $\eta_{t}= \rho C_{\nu}l_{\Delta} q$ is the dynamic turbulent viscosity and
\begin{equation}
S^{*}_{ij} = {1 \over 2} \left({\partial v_{i} \over \partial x_{j}}  + {\partial \nu_{j} \over \partial x_{i}} \right) 
- {1 \over 3} \delta_{ij}{\partial v_{i} \over \partial x_{i}}
\end{equation}
is the trace-free rate-of-strain tensor. Further details and numerical validations of the above closures are given in dedicated studies \citep{SchmNie06c,2009ApJ...707...40M}.

For our SGS model, the coefficients are calibrated against transonic compressible turbulence simulations, comparable to the regime in our simulations \cite{SchmNie06b}. The method of adaptively refined large eddy simulations is used to apply the SGS model in cosmological AMR simulations \citep{2009ApJ...707...40M}. As the SGS turbulence energy depends on the grid scale, which varies on adaptive meshes, energy must be exchanged with the numerically resolved velocity field when the grid is refined or de-refined. This is achieved by assuming the Kolmogorov two-thirds law for the scaling of the turbulent velocity fluctuations. As long as the turbulence is subsonic and nearly isotropic locally, this is a reasonable assumption. In contrast, the method used in \cite{2011ApJ...733...88G} does not calculate the turbulent energy on the grid scale, but the energy associated with a characteristic length scale of buoyant bubbles. Our SGS model completely neglects gravity on unresolved length scales and assumes the turbulent cascade from larger resolved scales as the dominant source of unresolved turbulence for the effects of buoyancy on subgrid scales \citep{SchmNie06c}. With a resolution of 64 cells per Jeans length, we expect this assumption to be valid on unresolved scales.

% The formation of black holes via direct collapse is feasible if fragmentation is sufficiently suppressed, so that most of the material is accreted onto the central object. For this purpose, we consider a primordial halo where molecular hydrogen is efficiently dissociated by radiative backgrounds, and cooling proceeds via atomic hydrogen lines \cite{Bromm03,2007MNRAS.374.1557J, 2010MNRAS.402.1249S, Schleicher10}. In order to transport the angular momentum, previous studies suggested scenarios based on gravitational instabilities in self-gravitating disks  \cite{2004MNRAS.354..292K} and self-regulation due to the presence of turbulence \cite{2009ApJ...702L...5B}. In a similar fashion, turbulence is thought to regulate the angular momentum transport in minihalos \cite{2002Sci...295...93A,2012ApJ...745..154T} and potentially cause fragmentation \cite{Clark11,Greif12}. For this purpose, we perform adaptively refined large eddy simulations, using the cosmological $N$-body and fluid dynamics code ENZO to resolve scales from $1$~Mpc down to the AU scale

Even though the presence of turbulence was reported in massive primordial halos already by \cite{2008ApJ...682..745W}, it was only realized by \cite{2011ApJ...731...62F} that a minimum resolution of $32$ cells per Jeans length is required to obtain converged turbulent energies during gravitational collapse. In addition, the turbulent energy cascade is expected to generate turbulence even on subgrid-scales (SGS), which may backreact on the resolved scales via turbulent viscosity. %WS
In engineering and many other disciplines of computational fluid dynamics, SGS turbulence models are used to represent the effect of unresolved turbulence on resolved scales in so-called large eddy simulations (LES). 
% By adopting this approach to cosmological adaptive mesh refinement simulations \citep{2009ApJ...707...40M}, these processes were recently shown to have significant impact in comparison to standard  simulations without an explicit SGS model \citep{2013MNRAS.tmp..551L}.

\section{Main Results}

\begin{figure*}
% \vspace{-4.0cm}
\centering
\begin{tabular}{c}
\begin{minipage}{6cm}
 \hspace{-3cm}
\includegraphics[scale=0.6]{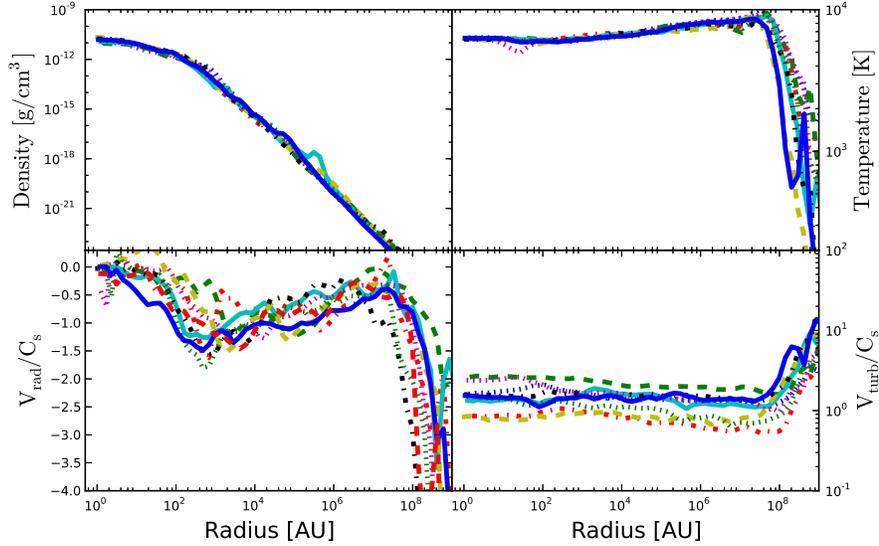}
\end{minipage}
\end{tabular}
\caption{The figure shows the radially binned spherically averaged radial profiles for nine different halos for SGS turbulence. Each line style and color represents a halo. The upper left panel of the figure shows the density radial profiles. The temperature radial profiles are depicted in the upper right panel. The bottom left panel of the figure shows the radial velocity profiles scaled by the sound speed. The averaged radial profiles of turbulent mach number ($\rm v_{turb}/C_{s}$, for the definition of $\rm v_{turb}$ see equation 4 of \citep{2013MNRAS.tmp.1155L} ) are depicted in the bottom right panel of the figure.}
\label{fig4}
\end{figure*}

\begin{figure*}
% \hspace{-4.0cm}
\centering
\begin{tabular}{c}
\begin{minipage}{6cm}
\includegraphics[scale=0.4]{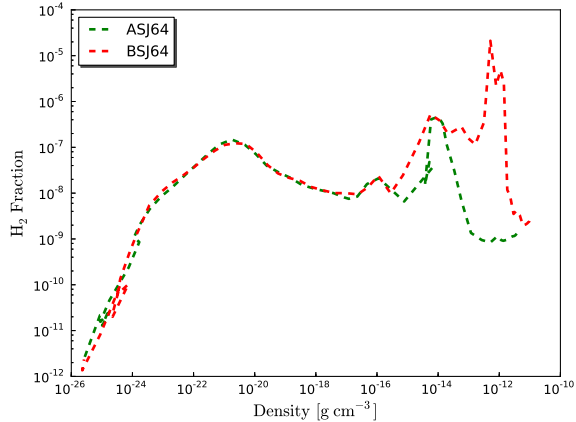}
\end{minipage}
\end{tabular}
\caption{The fraction of molecular hydrogen for two representative cases is shown in the figure. It can be found that this fraction remains lower than universal abundance of $\rm H_{2}$.}
\label{figh2}
\end{figure*}

% \begin{figure*}
% \hspace{-9.0cm}
% \centering
% \begin{tabular}{c}
% \begin{minipage}{4cm}
% \includegraphics[scale=0.6]{Turbprod1.ps}
% \end{minipage}
% \end{tabular}
% \caption{The figure shows the ratio of turbulence production to dissipation for various haloes.}
% \label{fig6}
% \end{figure*}

\begin{figure*}
\hspace{-9.0cm}
\centering
% \begin{tabular}{c}
% \begin{minipage}{4cm}
\hspace{9cm}
\includegraphics[scale=0.2]{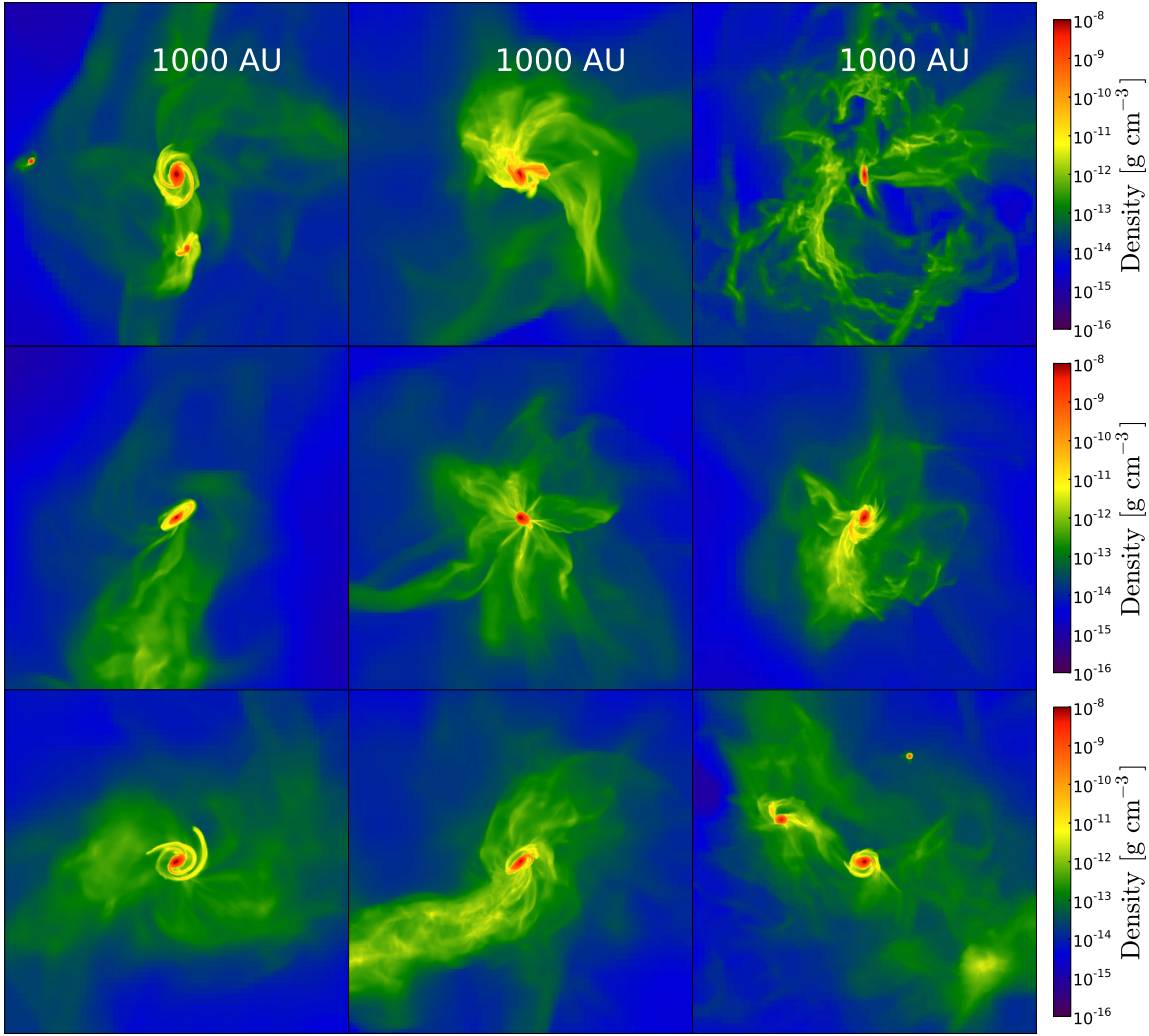}
% \end{minipage}
% \end{tabular}
\caption{The figure illustrates the state of simulations at the collapse redshifts for 9 halos. Density projections are shown for the central 1000 AU of the halo for a fixed Jeans resolution of 64 cells.}
\label{fig1}
\end{figure*}

\begin{figure*}
\hspace{-9.0cm}
\centering
% \begin{tabular}{c}
% \begin{minipage}{4cm}
\hspace{9cm}
\includegraphics[scale=0.2]{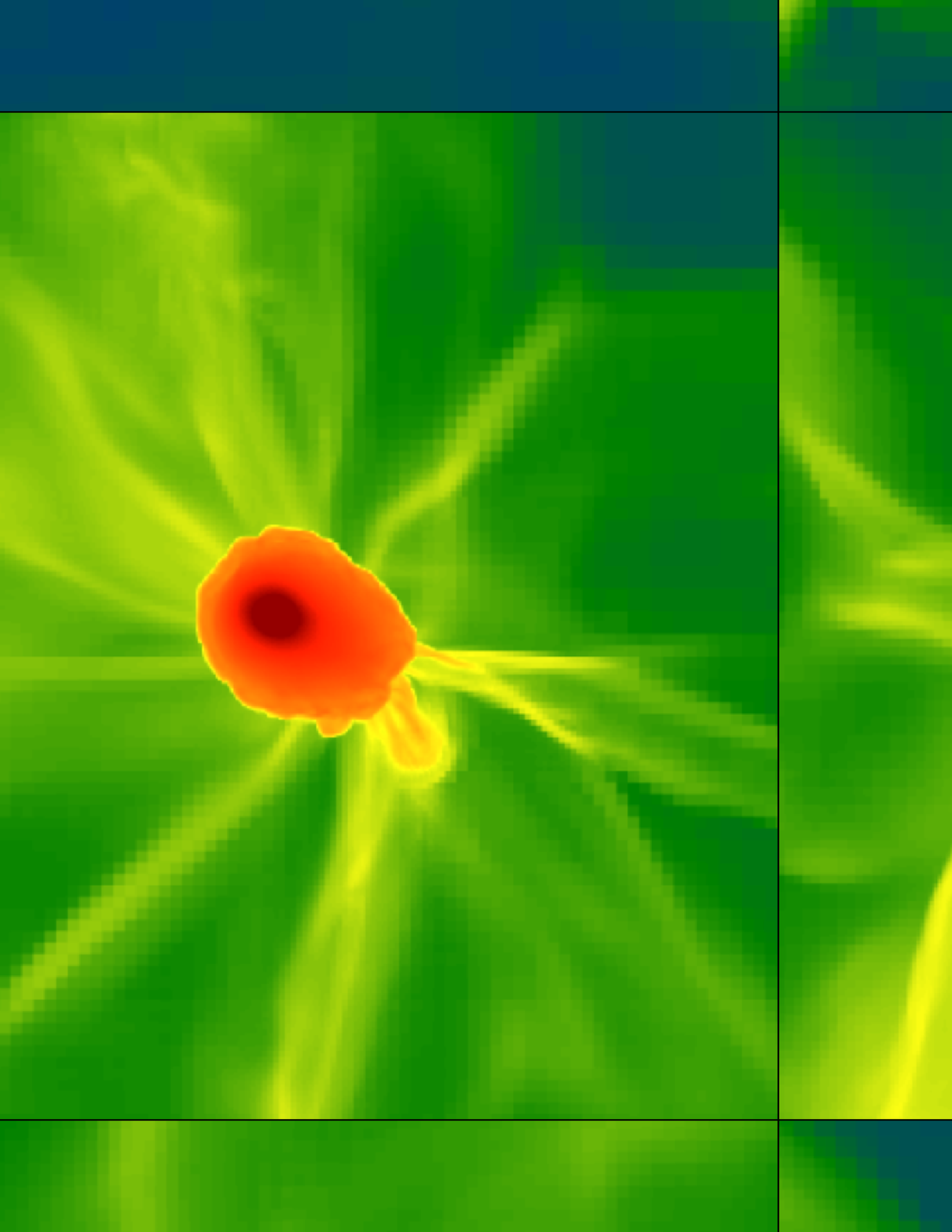}
% \end{minipage}
% \end{tabular}
\caption{Density projections for nine halos (same as figure \ref{fig1}) in the central 300 AU.}
\label{fig2}
\end{figure*}

In total, we have performed 9 LES and 9 equivalent ILES for nine halos as listed in table \ref{table1} for a constant radiation background with $\rm J_{21}=10^{3}$ of the $\rm H_{2}$ photo-dissociating radiation field for a radiation temperature of $\rm 10^{5}$ K. The results obtained from our cosmological simulations are presented here in detail. In the early phases of the collapse gas falls into the dark matter potential and gets shock-heated during the nonlinear evolution phase. Gravitational energy of the halo is continuously transferred to kinetic energy of the gas and dark matter during the course of virialization. 

\subsection{LES Runs}
The properties of the halos at larger scales during the formation of the first peak are shown in figure \ref{fig4}. The density profiles follow an $\rm R^{-2}$ behavior for all halos  according to the expectation of an isothermal collapse. The small bump in the density profile of one halo shows the formation of an additional clump. The top right panel of figure \ref{fig4} shows the temperature radial profiles. The gas in these halos is heated up to their virial temperatures and subsequently cools via Lyman alpha radiation. It is found that all halos collapse isothermally in the ubiquity of strong UV photo-dissociation flux and the formation of $\rm H_{\rm 2}$ remains suppressed. The temperature in the center of the halos is about 8000 K. The dissimilarities in temperature profiles around radii of $10^{9}$ AU arise from the difference in halo masses. The radial profiles of radial velocity scaled by the sound speed are depicted in the bottom left panel of figure \ref{fig4}. Overall, this ratio is negative which indicates the infall of gas to the center of the halo. This behavior is observed for all the halos. The turbulent Mach number is plotted in the bottom right panel of the figure \ref{fig4}. The value of turbulent Mach number is about one which indicates that turbulence is trans-sonic in the haloes. We attribute small variations in the turbulent Mach number to the differences in halo masses. The fraction of molecular hydrogen is shown in figure \ref{figh2} for a few representative cases. The H$_{\rm 2}$ fraction increases at lower densities due to the rise in electron abundance during the non-linear phase of the collapse. At intermediate densities, the H$_{2}$ abundance becomes constant as gas cools, recombines and remains neutral with a constant temperature around 8000 K. The presence of sharp spikes in the H$_{\rm 2}$ fraction is due to the shocks occurring at the central densities due to collisional dissociation as well as due to the presence of clumps. In general, the H$_{\rm 2}$ fraction is lower than the universal value (i.e., $\rm 10^{-3}$). Therefore, the contribution of H$_{\rm 2}$ cooling in the thermal evolution of the halos studied here is negligible. The overall halo properties are in accordance with previous studies \citep{2002Sci...295...93A,2008ApJ...682..745W,2009Sci...325..601T,2012ApJ...745..154T}.

% Probing the direct collapse scenario requires to follow the evolution of the halo beyond the initial collapse. Here we present the first high-resolution simulations following the evolution of the halo beyond the formation of the first peak, thus probing the subsequent accretion and fragmentation of the halo. %WS
% During the collapse, we ensure a numerical resolution of at least $64$ cells per Jeans length in order to resolve turbulent structures, and employ a SGS model for turbulence on unresolved scales \cite{2009ApJ...707...40M}. We further incorporate the effects of primordial chemistry in a strong photo-dissociating background of strength $\rm J_{21}=1000$ where $\rm J_{21}$ is the flux below the Lyman limit (i.e. 13.6 eV) in units of $\rm 10^{-21}~erg~cm^{-2}~s^{-1}~Hz^{-1}~sr^{-1}$. To follow the evolution beyond the formation of the first peak, we turn to an adiabatic evolution at a density of $10^{-10}$ g~cm$^{-3}$, leading to the formation of an adiabatic core. We continued the further evolution until reaching a core density of $\rm 1.2\times10^{-8}$~g~cm$^{-3}$, corresponding to at least $25$ years of further evolution or four free-fall times. Such simulations have been performed for a total of 9 halos, in order to investigate the potential differences from halo to halo (see table 1). For comparison, a corresponding set of 9 simulations has been performed without the SGS model, which are presented in the supplementary material, along with an additional description of the methods employed here. 

As a result of the strong photodissociating background, the halos go through an almost isothermal collapse regulated by atomic hydrogen line cooling. The central core develops a highly turbulent, extended structure. When the highest refinement level is reached, the evolution becomes adiabatic, thus stabilizing the gas on the finest grid and preventing collapse to smaller scales. The evolution on larger scales however continues, leading to turbulent accretion onto the central clump and occasional fragmentation as seen in Fig.~\ref{fig1}. In the final stage of the simulation, fragmentation is clearly visible in 3 out of 9 halos. Even in some of the other halos, fragmentation occasionally occurs, but the clumps merge again on short timescales. Zooming in further, we note that the central clumps are surrounded by self-gravitating accretion disks with prominent spiral arm structures depicted in Fig.~\ref{fig2}. The growth of the clumps is thus regulated due to accretion via the observed spiral arm structures as well as occasional mergers.

\begin{figure*}
\hspace{-4.0cm}
\centering
\begin{tabular}{c}
% \begin{minipage}{6cm}
% \includegraphics[scale=0.50]{MassDens_profile.ps}
% \end{minipage} &
% \hspace{3.8cm}
\begin{minipage}{8cm}
\hspace{-2cm}
\includegraphics[scale=0.70]{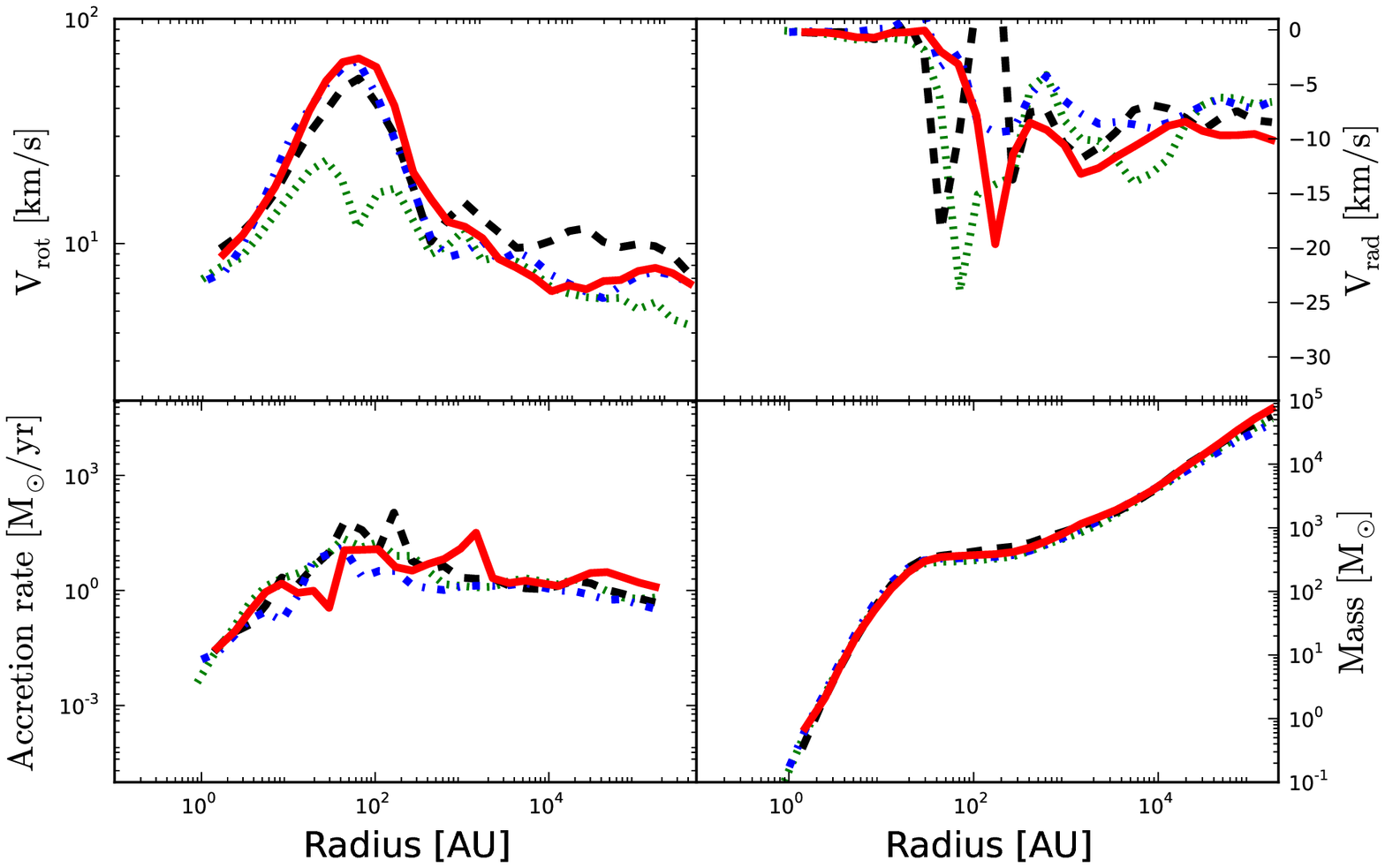}
\end{minipage}
\end{tabular}
\caption{Radial profiles for the physical properties of the disks centered on the peak density at collapse redshifts. The rotational and radial velocity profiles are shown in the top panels. Mass accretion rate ($4\pi R^{2}\rho v_{rad}$) and mass radial profiles are shown in the bottom panel. These profiles are shown for 4 representative halos.}
\label{fig0}
\end{figure*}

\begin{figure*}
\hspace{-2.0cm}
\centering
\begin{tabular}{c c}
\begin{minipage}{6cm}
\includegraphics[scale=0.4]{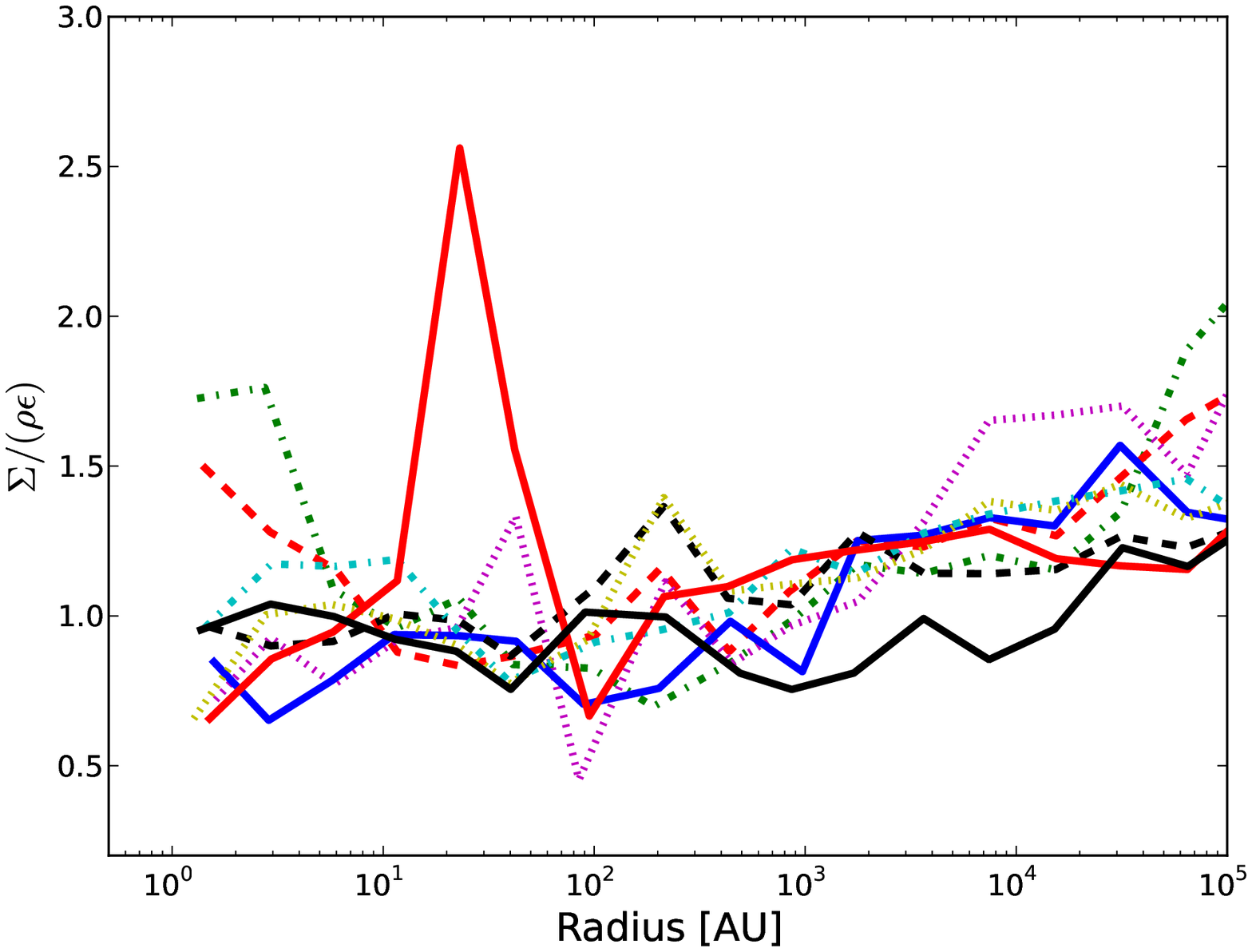}
\end{minipage} &
\hspace{1cm}
\begin{minipage}{6cm}
\includegraphics[scale=0.4]{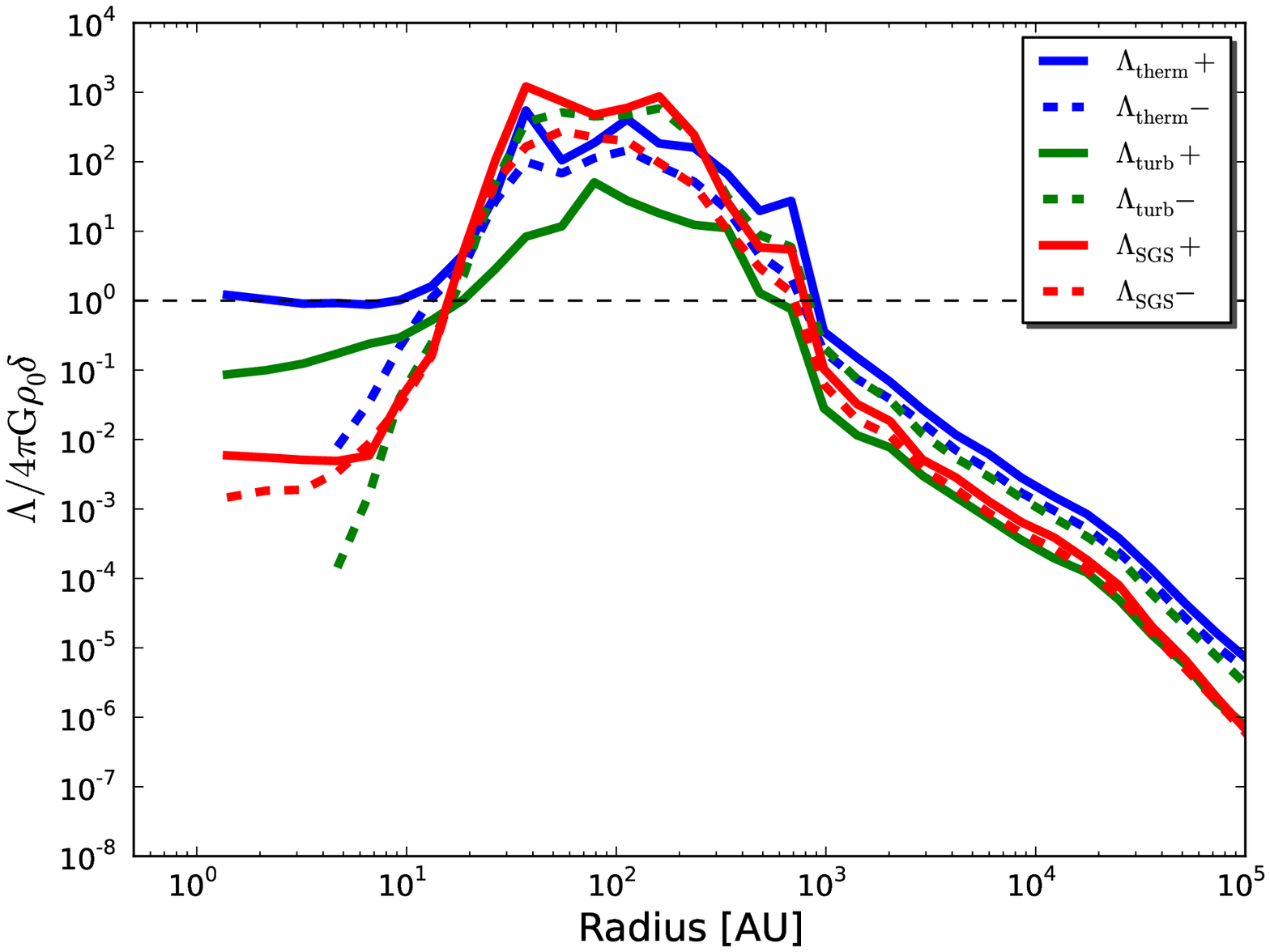}
\end{minipage}
\end{tabular}
\caption{The left panel of this figure shows the ratio of turbulence production to dissipation for various haloes at their collapse redshifts. Comparison of local support terms like thermal pressure, resolved turbulence and SGS turbulence for a representative case is shown in the right panel. They are scaled by the gravitational compression. The positive components mean support against gravity while negative components aid the compression.}
\label{fig3}
\end{figure*}

\begin{figure*}
\hspace{-2.0cm}
\centering
\begin{tabular}{c}
\begin{minipage}{8cm}
\includegraphics[scale=0.5]{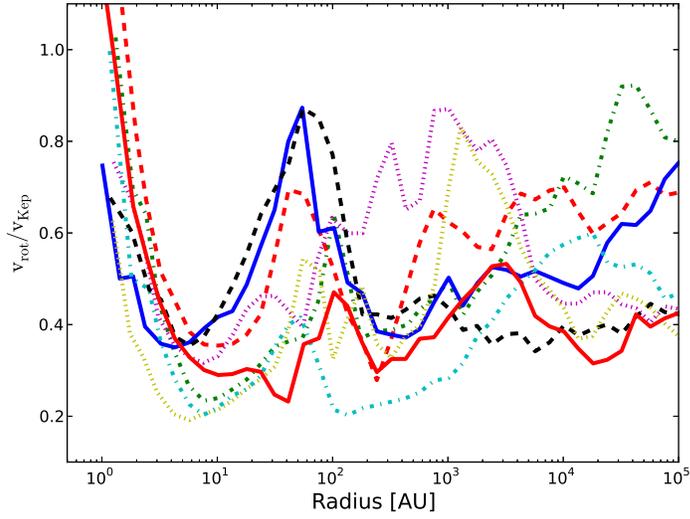}
\end{minipage}
\end{tabular}
\caption{The ratio of rotational to Keplerian velocity for the disks shown in figure \ref{fig2}. Each line style and color represents one disk.}
\label{fig5}
\end{figure*}

\begin{figure*}
\hspace{-9.0cm}
\centering
% \begin{tabular}{c}
% \begin{minipage}{4cm}
\hspace{9cm}
\includegraphics[scale=0.2]{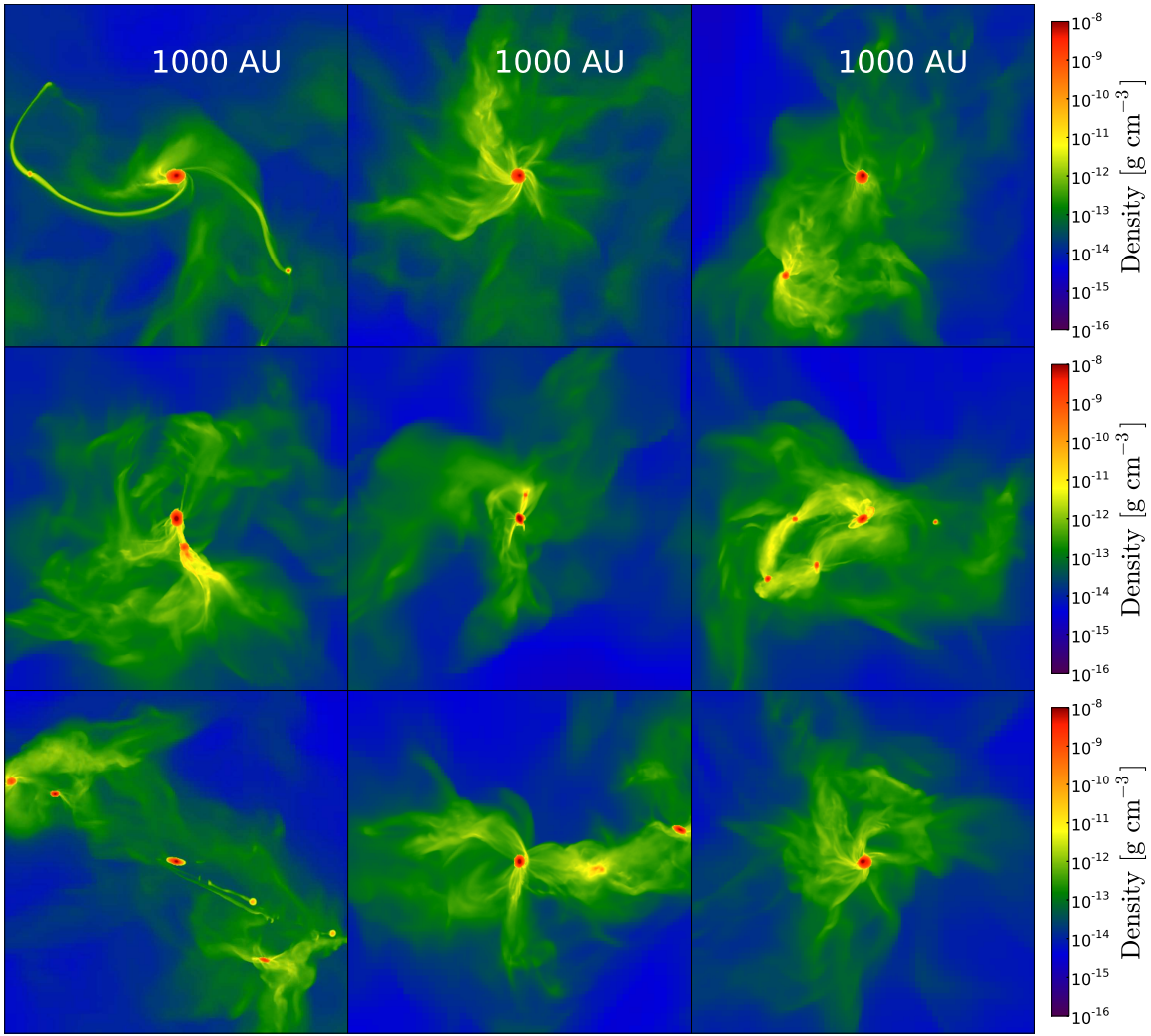}
% \end{minipage}
% \end{tabular}
\caption{The figure illustrates the state of simulations for various halos (ILES) at their collapse redshifts (at $t_{ff}=4$). The density projections are shown for the central 1000 AU of the halos.}
\label{fig8}
\end{figure*}

\begin{figure*}
\hspace{-2.0cm}
\centering
\begin{tabular}{c}
\begin{minipage}{8cm}
\includegraphics[scale=0.5]{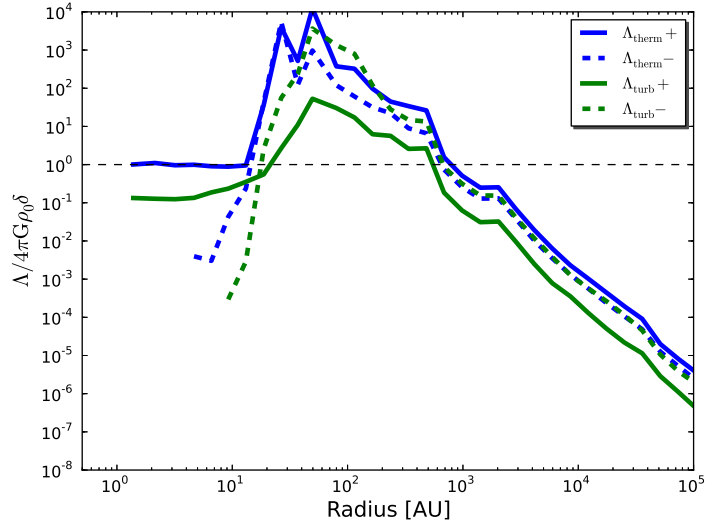}
\end{minipage}
\end{tabular}
\caption{The local support terms for a representative case are shown at collapse redshift (at $t_{ff}=4$). The solid lines represent the positive component while dashed lines indicate the negative component of the support term. The positive components mean support against gravity while negative components aid the compression}.
\label{fig7}
\end{figure*}

\begin{figure*}
\hspace{-9.0cm}
\centering
% \begin{tabular}{c}
% \begin{minipage}{4cm}
\hspace{9cm}
\includegraphics[scale=0.2]{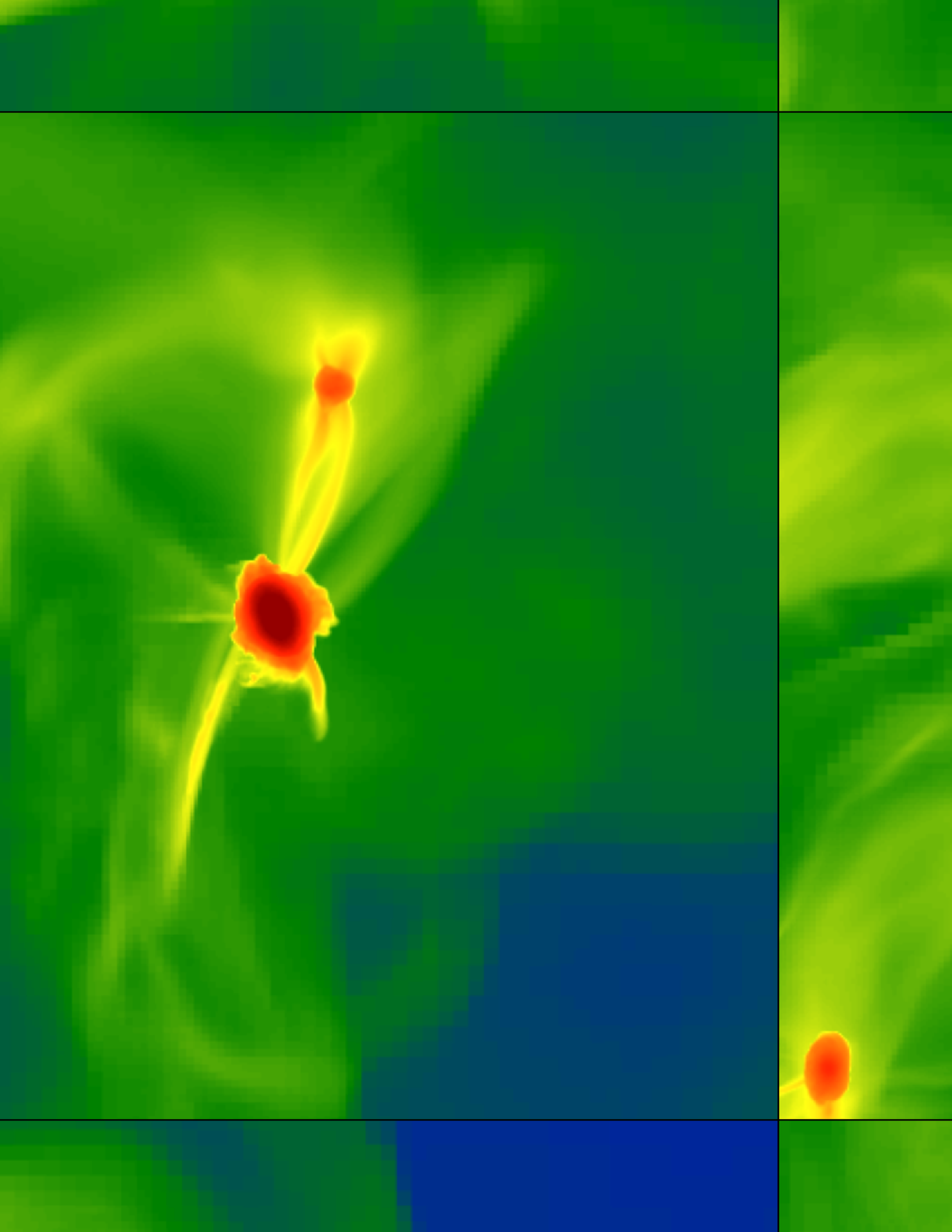}
% \end{minipage}
% \end{tabular}
\caption{Same as figure \ref{fig8}, density projections for the fields of view of 300 AU, for the ILES runs. }
\label{fig9}
\end{figure*}

The radial profiles of the physical properties of the halo in the central $10^{5}$ AU are shown for four representative cases in figure \ref{fig0}. The mass profile increases as $\rm R^3$ in the region of constant density, is then approximately constant within the disk, thus implying a minor contribution of the disk to the total mass, and subsequently increases as $\rm R$ according to the density profile. The latter shows $\rm R^{-2}$ behavior corresponding to an isothermal collapse, while flattening occurs in the central clump, where the collapse is stabilized by an adiabatic core. The rotational velocity ($\rm V_{rot}= v \times {\hat{a} \over |a|}$ where $ \rm \hat{a}= L \times r$, L is the specific angular momentum and r is the radius ) is about 10 $\rm km/s$ in the center, peaks at radii of 100 AU and then declines as it follows the Keplerian rotation. Variations in the peak rotational velocity arise due to the difference in the halo mass and spin. The radial velocity profile indicates that gas is falling into the center of the halos with typical velocity of about 10 $\rm km/s$. The mean accretion rate for these halos is thus a few $\rm M_{\odot}/yr$.

\textbf{Turbulence production and dissipation}:
We found that the medium inside the halo becomes highly turbulent and clumpy due to well resolved turbulence below a scale of $\rm \sim 10^{5}$ AU. The turbulent energy is low in the outskirts of the halo and tends to be stronger in the core. The rate of SGS turbulence production is computed as
\begin{equation}
\Sigma= -\tau_{ij} {\partial v_{i} \over \partial x_{j}},
\end{equation}
while the turbulent dissipation rate $\rho \epsilon$ is defined in equation~(\ref{edis}). The ratio of turbulence production to dissipation for different LES is depicted in  the left panel of figure \ref{fig3}.  In the outer regions, the turbulent production rate tends to be higher than the dissipation rate. This is a consequence of gravitational instabilities exciting turbulent motions during the collapse. In the adiabatic cores, on the other hand, production and dissipation are roughly in equilibrium, which indicates developed turbulence. However, local enhancement of turbulence production by instabilities can be seen even in the inner regions of some halos. 

On the basis of a simple Kolmogorov scaling for homogenous turbulence, one would naively expect the turbulent energy to decrease towards small radii. This effect is compensated as turbulence is produced locally during the gravitational collapse \cite[see also][]{2013MNRAS.tmp..551L}. In order to elucidate the contribution of turbulence, we followed the dynamics of gas compression employing a differential equation for the divergence of the velocity field $d=\nabla \cdot v$ \citep{2013MNRAS.tmp.1070S}:
\begin{equation}
- {D d \over Dt}= 4 \pi G \rho_{0}\delta -\Lambda
\end{equation}
%WS\[- {D d \over Dt}= 4 \pi G \rho_{0}\delta -\Lambda\]
Here, ${D \over Dt}= {\partial \over \partial t} + v \cdot \nabla$, $\delta$ is the overdensity relative to the mean density $\rho_0$ and 
$\Lambda$ the local support terms against gravitational compression. $\Lambda$ receives the contributions from thermal pressure, resolved turbulence and the SGS turbulent pressure.

\textbf{Support terms :} 
Thermal pressure support against gravity defines the critical mass for gravitational instability to become effective and is often accounted using the Jeans mass. However, in a strict theoretical sense, it is not applicable to substructures in a turbulent self-gravitating medium. For a more sophisticated analysis, we have computed the local support by thermal pressure, turbulence on numerically resolved scales and SGS turbulent pressure. These local support terms are denoted by $\Lambda_{therm}$, $\Lambda_{turb}$ and  $\Lambda_{SGS}$. They are defined as \citep{2013MNRAS.tmp.1070S}:

\begin{equation}
\Lambda_{\rm therm} = -{1 \over \rho} {\partial^{2}P \over \partial x_{i} \partial x_{i}} + {1 \over \rho^{2}} {\partial \rho \over \partial x_{i}}{\partial P \over \partial x_{i}}
\end{equation}

\begin{equation}
\Lambda_{\rm turb} = {1 \over 2} \left( \omega^{2}- |S|^{2} \right)
\end{equation}
where P is pressure, $\rho$ is the gas density, $\omega = \nabla \times v $ is the vorticity of the fluid and $|S| = (2 S_{ij}S_{ij})^{1/2}$ is the rate of strain

\begin{equation}
S_{ij} = {1 \over 2} \left( { \partial v_{i} \over \partial x_{j}} + { \partial v_{j} \over \partial x_{i}} \right)
\end{equation}

\begin{equation}
\Lambda_{\rm SGS} = -{1 \over \rho} {\partial^{2}P_{SGS} \over \partial x_{i} \partial x_{i}} + {1 \over \rho^{2}} {\partial \rho \over \partial x_{i}}{\partial P_{SGS} \over \partial x_{i}}
\end{equation}
Here $P_{\rm SGS}=\frac{2}{3}\rho e_{\rm t}$ is the SGS turbulent pressure. If these source terms are positive they provide support against gravity otherwise they aid to gas compression. These terms regulate the evolution of the divergence of the velocity field given by
\begin{equation}
- {D d \over Dt}= 4 \pi G \rho_{0}\delta -\Lambda.
\end{equation}
% It is found that the net support by SGS turbulence becomes relevant around radii of 100 AU and acts against the gravitational compression. Hence, it leads to the formation of disks as discussed in the main manuscript. 
%  

The positive and negative contributions of these terms scaled by the gravitational compression rate $4 \pi G \rho_{0}\delta$ are plotted in figure \ref{fig3}.
The plot clearly demonstrates that dynamical compression by gravity dominates above 1000 AU because $\Lambda/4 \pi G \rho_{0}\delta \ll 1$.
The thermal support of the gas becomes dominant and acts against gravitational compression at radii smaller than 10 AU, which corresponds to the center of our adiabatic core.
In between these scales, while the resolved velocity field compresses the gas (negative net support by resolved turbulence), this effect is balanced by the large positive support from SGS turbulence. In this region, self-gravitating disks are able to form due to the additional turbulent viscosity. Thus, turbulence has both a stabilizing and compressive effect, where the former is mainly due to eddies on length scales below the resolution limit. The positive support by resolved turbulence may also include the contribution from rotation. However, it is important to note that later is subdominant i.e., the main positive support results from SGS model, while resolved turbulence appears to have predominantly compressive modes.

A further analysis of the central regions shows that the disks seen in Fig.~\ref{fig2} are rotationally supported and stable against gravitational instability. The ratio of rotational to Keplerian velocity for disks is shown in figure \ref{fig5}. It is found that the disks are partly rotationally supported as the mean rotational velocity is 0.4 times the Keplerian velocity.

\subsection{ILES Runs}

The general properties of the halos for standard hydro runs without SGS model are identical to those found from the LES (see also \cite{2013MNRAS.tmp..551L}). All halos collapse isothermally and follow  an $\rm R^{-2}$ behavior. The central temperature in the halos before the adiabatic evolution is about 8000 K. No differences are observed in the thermal properties of the halo. 

We however see strong differences in the morphology and the fragmentation of the halos compared to the LES. The state of the simulations on a scale of 1000 AU is shown in figure \ref{fig8}. It is found that fragmentation is more common in these runs (7 out of 9 halos). Nevertheless, we see that a massive object is formed also in the core of these halos. The typical masses of these clumps are slightly smaller than in the LES in most of the runs due to the presence of additional clumps. The masses of the most massive clumps are listed in table \ref{table1}. The formation of less massive clumps is much more common in these runs. We attribute the formation of clumps to turbulent compression which locally aids in gravitational compression and helps in the formation of smaller clumps. Here this effect is enhanced as the viscous damping from unresolved turbulence is missing in these calculations. 
To quantify the effect in more detail, we have compared the contribution of the local support terms. They are shown in figure \ref{fig7} for a representative case. It can be noticed that gravitational compression dominates at larger scales while the resolved turbulence becomes important in the central 100 AU, where the latter becomes comparable to the support from thermal pressure. It may even exceed the thermal support locally. Overall, the support from resolved turbulence is negative which aids in gravitational compression. In the very center of the halo, we see an adiabatic core which is supported by thermal pressure.

We also found that in some cases these clumps are produced at earlier times and merge with the central object. It can be noticed from the figure that these clumps are very well separated and may lead to the formation of binary or multiple systems. Depending on the further evolution the clumps might evolve to a massive star cluster surrounding a central black hole. We  see accretion rates of few $M_{\odot}/yr$ comparable to LES. Zooming into the central object as shown in figure \ref{fig9}, the clumps are dense and compact. No disks are observed in these runs due to the higher turbulent velocities. Nevertheless, an object of mass between 500-1000 $\rm M_{\odot}$ forms in the center of every halo which will lead to formation of massive black hole or a supermassive star as an intermediate stage.
% The complex interplay between turbulent stresses and gravitational compression leads to the formation of massive clumps and is highly nonlinear process.

\begin{table*}
\begin{center}
\caption{Properties of the simulated halos are listed here.}
\begin{tabular}{cccccc}
\hline
\hline

Model	& Mass			& spin parameter     & Collapse redshift    & Clump Masses & Clump Masses\\

 & $\rm M_{\odot} $	 & $\lambda$    & z  & LES ($\rm M_{\odot}$) & ILES ($\rm M_{\odot}$)\\ 
\hline                                                          \\
 A	  & $\rm 8.06 \times 10^{6}$	& 0.0347468	&12.06 &950 &460\\		
 B	  & $\rm 4.3 \times 10^{6}$	& 0.0309765 	&11.3  &850 &850\\
%  C	  & $\rm 3.2 \times 10^{7}$	& 0.0310164	&14.0  &   &12.06\\	
 C        & $\rm 2 \times 10^{7}$       & 0.0178532     &12.6  &800  &611 \\
 D        & $\rm 1.0 \times 10^{7}$    & 0.0338661      &12.8  &850  &842\\
 E        & $\rm 1.9 \times 10^{7}$     & 0.0084786     &13.7  &1200 &741\\
 F        & $\rm 4.5 \times 10^{7}$    & 0.0294066      &18.1  &800 &588\\
 G        & $\rm 2.3 \times 10^{7}$     & 0.021782      &15.9  &800 &815\\
 H        & $\rm 9.7 \times 10^{6}$     & 0.0099387     &13.5  &900 &1522\\
 I        & $\rm 8.2 \times 10^{6}$     & 0.0252206     &15.0  &556 &1000\\
\hline
\end{tabular}
\label{table1}
\end{center}

\end{table*}

\section{Discussion}

We present here the highest resolution cosmological large-eddy simulations to date which track the evolution of high-density regions on scales of $0.25$~AU beyond the formation of the first peak, and study the impact of subgrid-scale turbulence. Our findings show that while fragmentation occasionally occurs, it does not prevent the growth of a central massive object resulting from turbulent accretion and occasional mergers. The central object reaches $\sim 1000$ solar masses within $4$ free-fall times. They may further grow up to $10^{6}$ solar masses through accretion in about 1 million years. The direct collapse model thus provides a viable pathway of forming high-mass black holes at early cosmic times.

The first massive clump is formed in the center of the halo as early as $\rm z=18.0$, and begins to accrete gas from the ambient medium. Due to the additional turbulent viscosity, the morphology of the halos becomes more extended compared to the runs without SGS model. At the final stage, our simulations harbor central massive objects with $\sim1000$~M$_\odot$. These objects are expected to form a supermassive star on the Kelvin-Helmholtz timescale, which may subsequently collapse to a supermassive black hole \citep{2001ApJ...550..372F}. In the presence of efficient accretion, the atmospheres of such massive protostars remain at low temperatures, and UV feedback is negligible \citep{2012ApJ...756...93H}. On the other hand, feedback via the accretion luminosity may become relevant at an earlier stage.
% Adopting the optically thin approximation, accretion heating may cause a temperature difference of at most 500 K at the end of our simulation, close to the massive central object. 

\textbf{Accretion Luminosity :}
We also computed the heating rate from the accretion luminosity for the massive clumps in the center of the disk employing the optically thin approximation, as
\begin{equation}
\Gamma_{acc} = \kappa_{P}\left({L_{acc} \over 4 \pi R^{2}}\right)~erg~g^{-1}~s^{-1}
\end{equation}
where $ \kappa_{\rm P}$ is the Planckian opacity, $R$ is the distance from the source and $\rm L_{\rm acc}$ is the accretion luminosity computed as
\begin{equation}
L_{acc} = {G M_{*} \dot{M} \over R_{*}}
\end{equation}
here $M_{*}$ is the mass of the star, $\dot{M}$ is the mass accretion rate and $R_{*}$ is the radius of star obtained from the relation \citep{2012ApJ...756...93H} 
\begin{equation}
R_{*} = 2.6 \times 10^{3} R_{\odot} \left({M_{*} \over 100 M_{\odot}}\right)^{1/2}
\end{equation}
Our estimates show that for a $500~M_{\odot}$ clump with a typical size of $R\sim100$~AU, a temperature of $7500$~K, a Planckian opacity of $0.04$ \citep{2005MNRAS.358..614M} and a typical accretion rate of $1$~M$_\odot$~yr$^{-1}$, the heating rate from the accretion luminosity is $2 \times 10^{-4}$~erg~cm$^{-3}$~s$^{-1}$, which is approximately equal to the atomic line cooling, given as $1.6 \times 10^{-4}$~erg~cm$^{-3}$~s$^{-1}$. Once the masses of the clumps reach 1000~$M_{\odot}$, the heating rate exceeds the cooling rate by a factor of $10$. We note, however, that this is easily compensated by a minor change in the temperature. Specifically, a temperature rise by $500$~K would be sufficient to account for the additional heating. However, with the optically thin approximation, we likely overestimate the contribution on the scales considered here. In fact, with an optical depth $\tau \sim R \rho \kappa_{\rm P} > 30$, the accretion luminosity feedback is likely confined to much smaller scales, thus not affecting the evolution considered here. It can thus be expected that the accretion luminosity heating would only cause minor differences until this point, and potentially stabilize the central region further, i.e., suppress fragmentation even further. 

We further note that this effect has been examined in further detail by \cite{2011MNRAS.414.3633S} for minihalos, finding that the amount of fragmentation is at most slightly reduced. We would expect that to favor the formation of a central massive object even further. However, in the atomic cooling halos considered here, atomic hydrogen is expected to act as an even more efficient thermostat, and the relative change of the thermal pressure even less relevant. 

% However, the gas is optically thick to such radiation, and the energy is therefore absorbed on much smaller scales, providing additional support against fragmentation. 

These massive clumps will collapse to similar-mass black holes at the end of their lifetime, which can grow to the observed $10^9$~M$_\odot$ black holes at $z\sim6-7$ via Eddington accretion \citep{Shapiro05}. The main requirement to obtain high-mass black hole seeds is thus a massive primordial halo at $z\sim15$, exposed to a sufficiently strong radiation field for H$_2$ dissociation. The abundance of such halos was previously calculated by \cite{2008MNRAS.391.1961D,2012MNRAS.425.2854A}, showing that these are indeed frequent enough to explain the number of observed high-redshift black holes. The high-resolution simulations presented here strongly confirm the feasibility of this scenario from a dynamical point of view, as a result of turbulent accretion and occasional mergers with additional fragments. For a final conclusion on the origin of supermassive black holes, future studies will need to address the growth of these clumps in the presence of feedback. At the later stages, we expect star formation and stellar feedback to affect the subsequent evolution, potentially giving rise to a co-evolution between stellar bulge and central black hole \citep{Tremaine02,2011MNRAS.410..919J}. The formation of their seeds is in any case feasible, thus yielding strong support for direct collapse scenarios.

\section*{Acknowledgments}
The simulations described in this work were performed using the Enzo code, developed by the Laboratory for Computational Astrophysics at the University of California in San Diego (http://lca.ucsd.edu). We acknowledge research funding by Deutsche Forschungsgemeinschaft (DFG) under grant SFB $\rm 963/1$, project A12 and computing time from HLRN under project nip00029. The simulation results are analyzed using the visualization toolkit for astrophysical data YT \citep{2011ApJS..192....9T}.

\bibliography{bibliobhs.bib}

\end{document}